\documentclass[12pt]{article}
\usepackage{graphicx}
\begin{document}

%
%
%

\title{On One Model of the  Geometrical Quintessence}

\author{\footnotesize V.GUROVICH\footnote{ Physics Institute of NAS, Chui av. 265a,
Bishkek, 720071, Kyrgyzstan; e-mail:astra@freenet.kg}\quad and
 I.TOKAREVA\footnote{Physics
Department, Technion, Technion-city, Haifa 32000, Israel; e-mail:
iya@tx.technion.ac.il}}
\date{}
 \maketitle



\begin{abstract}
A cosmological model with modification of  the Einstein-Hilbert
action by the correction $f(R)\propto \beta R^n$ is considered.
Such way of the description of the ``geometrical'' dark energy has
been introduced repeatedly and the  coefficients of the model were
chosen to be fit against some observational data. In this paper
the unambiguous choice of parameters $n$ and $\beta$ is proposed
from the follow reasons:  the exponent $n$ close to $1.2953$
follows from the request for the evolution of the Universe after
recombination to be close to the evolution of the flat FRW model
with cold dark matter and the reasonable age of the Universe
defines the magnitude of the coefficient $\beta$. Such a model
corresponding to the evolution of the Universe with the dynamical
$\Lambda$-term   describes well enough  the observational data.
 \end{abstract}

\section{Introduction}
The discovery of accelerated expansion of the
 Universe \cite{Riess,Perlmutter,Tegmark} has stimulated the quest
  for mechanism of the modern inflation.
  The most famous theoretical model of  dark energy (DE) is the
cosmological constant $\Lambda$. The corresponding FRW
 solution for flat Universe with the present densities ratio for
  cold matter and dark energy ($\Omega_{m}/\Omega_{\Lambda0}
  \sim 0.3/0.7$) describes satisfactorily the evolution of the Universe
   at
low redshifts~\cite{Tegmark,Linde,sahni}. However,  the nature of
the
 constant $\Lambda$- term has been remaining to be inexplicable
 during many years.

It is well-known,  the application of the constant $\Lambda$- term
for  modeling of the early Universe was initially confronted with
principal difficulties. Solving  the problems of  the very early
Universe, this term has to be reducing by several order of
magnitude during following  evolution of the Universe. This
problem was solved by rejection of $\Lambda$- term,
  and corresponding inflationary behavior was determined by models
  of `` effective $\Lambda$- term''- quasi-classical scalar fields
  in the one way.  The progress of these
   models is well-known.
   Another way to describe the inflationary behavior
   is to take into account the polarization of
 vacuum of quantum fields in the  early Universe. Taking into
 account of the effects of  the polarization leads to  the
 appearance of the  terms non-linear on curvature in the
 Einstein-Hilbert
 action. In such models  the inflation appears
 self-consistently. Let us note two issues\\
- a correction to the Einstein -Hilbert  action  with the
arbitrary function of scalar curvature $R$
  is equivalent  mathematically to the
  introduction of scalar field into the classical Friedman
  cosmology\cite{Barrow,Baib,Star};\\
  - the terms of the form $R^n$ were investigated in
   the early works done on the problem of singularity before
  obtaining of exact corrections to Einstein action following
  from the one loop approximation
  \cite{gur1}. The part
  of such solutions approaches asymptotically to Friedman solutions
  with $\Lambda$-term, however physical results of the solutions
  have   not been explained at that time.

   For the purpose to explain
   the  accelerated expansion of  the Universe today, it is
   naturally to use
   the experience
  cumulative in investigations  of early Universe. Thus,  one of
  the tendencies is concerned with the hypothesis of existence
  some scalar fields that determine the density of dark energy
  (DE) (e.g. \cite{Ratra,Zlatev,Starobinsky}). The another  tendency models an effective
  quasi-hydrodynamical tensor of momentum-energy
  describing the observational data \cite{kamen}. And third tendency
   is to
  generalize the Einstein-Hilbert equation by inclusion of
   curvature invariants
  \cite{Turner,Capozziello,fol} analogously to  the more early works
  mentioned above.
  The last approach one can consider  to be
   either  an  independent approach to the describing of DE or  an
 analogue of inclusion of scalar fields (in the case $f(R)$)
  as stated above.

 In the works  on the higher order gravity theories (HOGT)
the models with power corrections were investigated, however they
have never been fitted to whole set of the observational data.

  In this paper, the model with correction $f(R)\propto R^n$ with
  $n>0$ is considered in detail for the purpose to correlate it with
 the observational data. In the other words, we would like to
 obtain the  model  that does not  conflict with the  scenario of
 the  large scale  structure formation (in past) and describes
 satisfactorily the Universe undergoing an accelerated expansion
 at present.  Therefore, at the minimum,
  in the framework of the $f(R)$-theories,
  we will obtain
 solutions  remind $\Lambda$CMD model  describing well enough by
 set of the observational data \cite{Tegmark,Linde}.
  However, as it has been mentioned by various authors (see
 \cite{sahni} and references therein) the observational data indicates
 the models of the dynamical DE. Hence, our second aim is  to search out
 such dynamical solutions in the framework of the
 HOGT and to find out whether  these solutions are
 preferable.

 This paper is built as follows. In the section II the basic
 equations of  the HOGT are presented. In the section III for the
 corrections of the form $f(R)\propto \beta R^n$ we find the exponent
 $n_1=1.2953$
  which allows the generalized Einstein equation for the scale factor
  $a$ to have a particular   solution corresponding to the flat
  FRW solution for cold  matter.
  We show that instability of the
  solutions that are close to this particular solution at $z>>1$
   may lead to  the accelerated behavior  of
  the model at present and the following asymptotic
    approach of the solution to
  the solution with the constant $\Lambda$- term.
   In the section IV we discuss our results and compare them  to
 observations. In the model there are only two free parameters of the
 model - the coefficient $\beta$ and a slight deviation of the parameter
  $n$  from $n_1=1.2953$ mentioned above.
   Fixed from one set of the observational
  data they  allows to obtain the rest of the set
  of the observational data.

\section{Basic equations}
   We will work in the matter frame with a spatially flat FRW
   metric
\begin{equation}
\label{metric}
ds^2=d\tau^2-a^2(\tau)(dx^2+dy^2+dz^2),
\end{equation}
with $a(\tau)$ being a scale factor. The non-dimensional
components of Ricci scalar and Ricci tensor are,
\begin{eqnarray}
\label{Ricci}
 R=-6\left(\frac{\ddot{a}}{a}+
\left(\frac{\dot{a}}{a}\right)^2\right),\quad
R_0^0=-3\left(\frac{\ddot{a}}{a}\right),\quad
\dot{a}=\frac{1}{H_0}\frac{da}{d\tau}.
\end{eqnarray}
Here the present Hubble parameter $H_0$ has been introduced to
turn to non-dimensional values. The equations that can be derived
by variation of the action
\begin{equation}
\label{action} S=\frac{M_P^2}{2}\int d^4x
\sqrt{-g}\left(R+f(R)\right)+\int d^4x\sqrt{-g}\,L_m
\end{equation}
are well-known for a long time \cite{Gur70}. It is useful to reduce
  its $0-0$ component , that is the third
order equation  in $a$, to  the second order
 equation  by introduction of a new function
\begin{equation}
\label{yR} y(a)=(\dot{a}{a})^2.
\end{equation}
Thus, Eqs.(\ref{Ricci}) are determined as
\begin{equation}
\label{Riccy}
  R=-3y'/a^2,\quad R_0^0=3(y-a\,y')/2,
  \quad y'=dy/da,
\end{equation}
and the observable parameters - the Hubble parameter $h$ and the
deceleration parameter $q$ - are determined by formulae
\begin{equation}
\label{obpar} h(a)=\sqrt{\frac{y}{a^4}},\quad
q=-\frac{\ddot{a}a}{(\dot{a})^2}=\left(1-\frac{d\ln(y)}
{d\ln(a^2)}\right).
\end{equation}
Further, the $0-0$ component of Einstein equation can  be
  represented in  the form
 \begin{equation}
 \label{eqY}
 y+\left[\frac{df}{dR}\left(y-\frac{a}{2}\frac{dy}{da}\right)
 -\frac{a^4}{6}\, f+a\,y\frac{d}{da}\left(\frac{df}{dR}
\right)\right]=\Omega_{m}\,a.
\end{equation}

The solution for the classical Friedman model with cold matter is
defined by $f(R)=0$ whereas for the model with  the constant
$\Lambda$-term it is defined by $f(R)=const$. Taken together they
describe $\Lambda$CDM model, so, according to
Eqs.(\ref{Riccy}),(\ref{eqY}) we have
\begin{equation}
\label{LCDM}
 y=\Omega_{m}\,a+\Omega_{\Lambda}\, a^4,\quad
\Omega_m=\rho_m/\rho_{c}, \quad
\Omega_{\Lambda}=\rho_{\Lambda}/\rho_{c},
\end{equation}
where $\rho_m$, $\rho_{\Lambda}$,  and $\rho_{c}=3 H_0^2/8\pi\,G$
are the cold matter density, the dark energy  density, and the
 critical density of FRW Universe at present, respectively. For
$4$-interval (\ref{metric}) $\Omega_m+\Omega_{\Lambda}=1$. The
present-day observational values  are close to
$\Omega_{m}=0.3,\quad\Omega_{\Lambda}=0.7$.

Notice that the important property of  Eq.(\ref{eqY})
is a possible asymptotic
exit of solutions to regime of an exponential expansion.
Indeed, according  to definition (\ref{LCDM}) this solution is
 $y=\Omega\, a^4$. In this case, the scalar curvature
approaches to constant $R_{\infty}\rightarrow-12 \Omega$. In the
case of a negligible effect of matter  this solution is a
 particular solution of Eq.(\ref{eqY}) at execution of
  the following condition
 \begin{equation}
 \label{rav}
    \Omega=\frac{f(R_{\infty})}{6(1-(df/dR)|_{\infty})}=const>0.
 \end{equation}

\section{Model with Power Corrections  in  Einstein -Hilbert
Action }

In this section we will investigate  corrections to the
 Einstein-Hilbert   action of the form
\begin{equation}
\label{fr}
 f(R)=-\alpha \,R^n.
  \end{equation}

 In such  a case, Eq. (\ref{yR}) can be
presented in the form
\begin{eqnarray}
\label{y}
  \beta\left[n(n-1)y''y+\frac{(1-n)}{2}(y')^2+n(4-3n)\frac{y'y}{a}
  \right]=\frac{(y')^{2-n}}{a^{4-3n}}(y-\Omega_{m}\,a),\nonumber\\
  \beta=(-3)^{n-1}\alpha.\qquad \qquad \qquad \qquad \qquad
\end{eqnarray}
A general approach to investigation of the last equation is given
in paper\cite{Gur79}.

 There are two parameters in the model, $\beta$ and $n$,
determined by observational data. In a number of papers these
parameters were chosen according to different
requirements~\cite{Capozziello,fol}. In the present paper we will
choose parameter $n$ from requirement of  closeness of evolution
of the our model to the classical solution for the  flat FRW
Universe with cold matter in the past. This fact  allows   this
scenario to be close to the scenario of the large scale structure
formation. This requirement can be realized on condition  the
classical Friedman solution
\begin{equation}
\label{dust}
  y=\Omega_{m}\,a
\end{equation}
presents a particular solution of Eq.(\ref{y}). It is easy to see
from Eq.(\ref{y}), the last condition is equivalent to the choice
of $n$ to be satisfying the equation
\begin{equation}
\label{eqn} n-1=2n(4-3n) \quad {\textrm{with roots}}\quad
n_1=1.295,\quad n_2=-0.129.
\end{equation}
The first of the roots leads to the type of models of
papers~\cite{Capozziello,fol}, while the second root corresponds
the models with correction of the form $\propto \mu/R^{|n_2|}$
investigated in the paper~\cite{Turner1}. As it will be shown
below, such a choice of $n$  approaches the model to the set of
the observational data in the best way.

\subsection{Behavior of the dust solution in
the $f(R)$-model in past}

After recombination, the evolution of the Universe has to be
described by the Friedman model with the cold matter. The
 dynamics of expansion is determined by stability of the
dust solution in the model (\ref{y}). If the solution is stable,
then the model evolves in the way  very close to classical one.
However,  the observational data at $z<1$ does not correspond to
such a scenario, i.e. we are interested in  dust-like solutions
which  are not stable in the model (\ref{y}) but the perturbations
do not grow catastrophically fast. Otherwise, the model does not
provides a sufficiently long period with $q\simeq0.5$ at $z\gg1$
required for the large structure formation.

For investigation of behavior of the solution (\ref{dust}) we will
search  a perturbed solution in the form
\begin{equation}
\label{pert1} y=\Omega\, a(1+\psi),\quad \psi\ll1.
\end{equation}
 In this linear approximation and close to the recombination time
 ($a\ll1$) Eq. (\ref{y}) yields,
 \begin{eqnarray}
\label{psi0}
  n\,a^2\,\psi''+a\,\psi'(2n-0.5)=0,
\end{eqnarray}
with the damping solution $\psi=C_1+C_2\,a^{(\frac{1}{2n}-1)}$.
Hence the solution (\ref{dust}) for $f(R)$-theory (\ref{fr}) with
$n=n_1$ satisfying Eq.(\ref{eqn}) asymptotically approaches to to
the flat FRW solution, i.e. it is stable and does not satisfies
the requirement stated above\footnote{\footnotesize It is
interesting to note that the mentioned exponents (\ref{eqn}) are
obtained in the recent paper~\cite{Multamaki} for the equation
equivalent to Eq.(\ref{y}). In $f(R)$ theory, the given exponents
allow obtaining of  the solutions for cold dark matter coinciding
with ones in the classical Friedman model of the universe. Also,
it  has been shown in the paper~\cite{Multamaki}  that these
solutions are stable in the framework of HOGT. Let us notice
 we slightly change exponent $n$ in the preset work to obtain
weakly unstable solutions adequate to the observational data.
Authors are grateful to authors of paper~\cite{Multamaki} kindly
attracting our attention to their results.}

As the next step on the path of  the choice of $n$,  we will look
for solutions of the $f(R)$ -theories (\ref{fr}) with $n$ which is
a little different from $n_1$,
\begin{equation}
\label{dn1}
  n=n_1+\delta n,\qquad  \delta n\ll 1.
\end{equation}
It is efficient  to rewrite  the condition (\ref{eqn}) in the form
\begin{equation}
\label{eqnd} \delta\,(n-1)=2n(4-3n),\quad \delta=1+\epsilon,\quad
\epsilon\ll1.
\end{equation}
\begin{figure}[pt]
\begin{center}
\includegraphics[width=140 mm,height=6.5 cm]{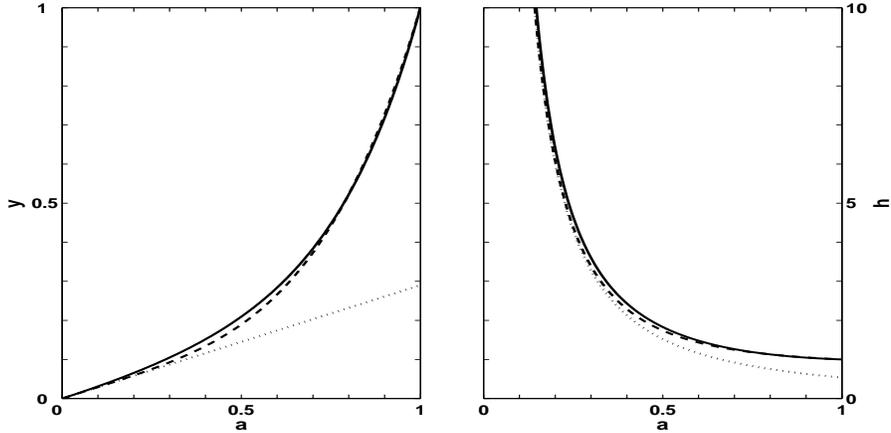}
 \caption{{\footnotesize The evolution of  the variable $y$ (a left
panel) and the Hubble parameter $h$ (a right panel) with the scale
factor $a$  are represented by the solid line. The $\Lambda$CDM
solution and the cold matter solution are presented by dashed and
dotted line, respectively. Both graphs  correspond to the case
$\Omega_{m}=0.29.$ \label{hy29}}}
\end{center}
\end{figure}
 We will look for the perturbed dust-like solution in the form
(\ref{pert1}) near to recombination. The Eq.(\ref{y}) for such a
case yields
\begin{eqnarray}
\label{psi1}
  n\,a^2\,\psi''+2a\,\psi'\left(n-0.5+\delta/4\right)
  +\left(\delta-1\right)=0.
\end{eqnarray}
Then after the change of variable  $\xi=\ln (a/a_{\ast})$ the Eq.
(\ref{psi1}) yields
\begin{equation}
 \label{psi1dot}
  n\ddot{\psi}+\left(n-1/2+\epsilon/2\right)
  \dot{\psi}+\epsilon=0,\,{\bf \dot{}}=d/d\xi.
\end{equation}

 The last equation has a solution
\begin{equation}
\label{pert}
    \psi=\left(C_1+\frac{2\,\epsilon}{1-2n}
    \ln\left(\frac{a}{a_{\ast}}\right)\right)+C_2\,a^{(\frac{1}{2n}-1)}.
\end{equation}
 The analysis of (\ref{pert}) has shown that the  requirement
 stated above is realized only for $\epsilon< 0$.

In this case, the modification of the exponent (\ref{eqnd}) is
determined by a small positive correction
\begin{equation}
\label{dn}
    \delta n \simeq -2\,\epsilon\, n_1(n_1-1)/(12n_1-1).
\end{equation}
 The numerical analysis has shown that behavior of solutions is
 sensitive to small changes of $\epsilon$ at $a\rightarrow1$. The
 last fact together  with a choice of parameter $\beta$ allows us
 to obtain a good enough  correspondence with the observational
data.

 As an illustration we  give  the results for the set of parameters
 $n=1.296,\; \beta=0.467$ fixed according to $\Omega_{m}=0.29, h_0=68$
  (the best fit to the CMB+SNe data presented in the paper~\cite{sahni}).
  In a left panel of Fig.~\ref{hy29},  a solid line represents
  the evolution of
the variable $y$  with a scale factor $a$. At the beginning,
 it coincides with  the evolution of the dust model
  which is represented by a
 dotted line but further it deviates to the $\Lambda$CMD model
   represented by
 a dashed line. In a right panel,
one can see evolution of the Hubble parameter with the scale factor $a$ for  the mentioned
three models.

\subsection{The behavior of the solution in future}
\begin{figure}[pt]
\begin{center}
\includegraphics[width=130 mm,height=6.5 cm]{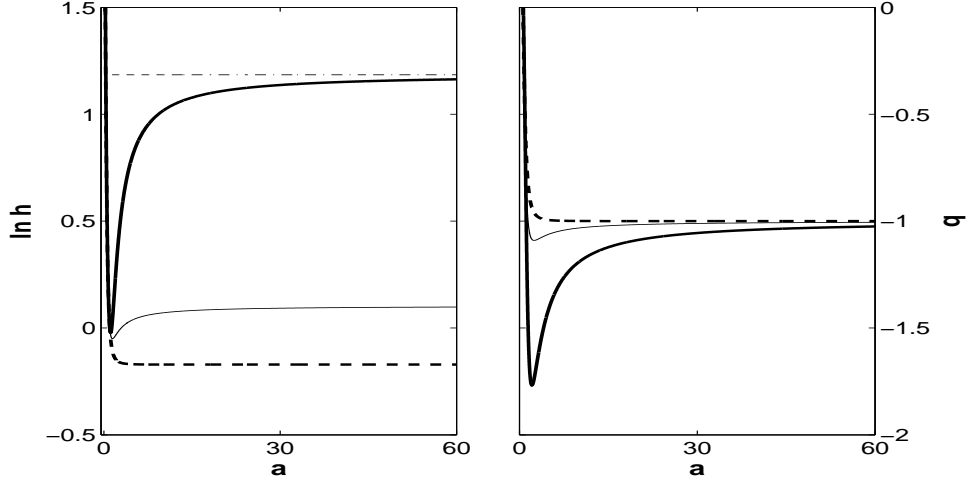}
 \caption{{\footnotesize The evolution of the Hubble parameter $h$
and the deceleration parameter $q$ with a scale factor $a$. The
thick solid lines represent graphs related to the $f(R)$-model
while the dashed lines represent  the evolution of the parameters
of $\Lambda$CDM model. In the left panel the asymptotic solution
for $f(R)$ model is shown by the dash-doted line.
 \label{future}}}
 \end{center}
\end{figure}
The further expansion  of the Universe at $a>>1$ according to the
Eq. (\ref{y}) leads to the negligible  effect of cold matter on the
solution behavior. In this case,   De Sitter
 solution $y=\Omega\, a^4$ is an asymptotic  solution
 of Eq.(\ref{y}). This solution corresponds to the
 constant Hubble parameter
 \begin{equation}
 \label{hOmega}
 h(a_{\longrightarrow\infty})=\sqrt{\Omega}
 \end{equation}
  with  $\Omega$ defined from equality
\begin{equation}
\label{Omega}
 3\beta(2n-1)(n-1)/2=(4\Omega)^{(1-n)}/4
\end{equation}

  The inflationary solution
 is stable in the
 process of evolution of  the model. To show it,
 we shall look for the solution with perturbation in the form\\
 $y=\Omega a^4(1+\Phi),\quad \Phi<<1$. This ansatz yields
\begin{equation}
\label{latepert}
 n_1a^2\Phi''+[(n_1-1)/2+3n_1^2]a\Phi' +6(2n_1^2-3n_1+1)\Phi=0
\end{equation}
The change of variable  $a$ to variable $\xi$ (see Eq.(\ref{pert}))
yields
\begin{equation}
\label{latepert2} \ddot{\Phi}+A\dot{\Phi} +B\Phi=0,\quad
\dot{}=d/d\xi.
\end{equation}
where  coefficients  are $A=[3n_1-(1/n_1+1)/2]=2.99$,
$B=6(2n_1-3+1/n_1)=2.18$. This equation for perturbations have a
damped  solution indicating De Sitter solution  to be
 stable. The numerical analysis has shown that De Sitter solution is an
attractive solution.

The evolution of the Hubble parameter is represented by the thick
solid line in the left panel of Fig.~\ref{future} for the case
$\Omega_m=0.29$. In contrast to
 the Hubble parameter of $\Lambda$CDM
model (the dashed line) monotonically decreasing down to constant
$\sqrt{\Omega_{\Lambda}}$,  it reaches  a minimum $h_m \approx
0.978$ at $a\approx 1.15$ and after that increases up to the
asymptotic solution (\ref{hOmega}) which is represented in the
figure by the dash-doted line. It is interesting to note that  the
formula for dimensionless parameter Hubble $h(z)=h(a)$
  obtained from observational data in the paper~\cite{sahni} allows
   it's extrapolation
in  future $(a>1)$. At the parameters  mentioned in this paper,
the formula for $h(a)$ also predicts minimum of the Hubble
parameter at  $a\approx 1.45$ which is equal to $0.951$.
 This best fit of the paper~\cite{sahni} is
shown in Fig.\ref{future} by the thin solid line.
  The deceleration parameter $q$
represented in the right panel of Fig.~\ref{future} also passes a
minimum and  approaches to $-1$ with the growth of $a$. Therefore,
we live in a transitional epoch between the classical Friedman
cosmology and the De Sitter cosmology.

 \section{Results}
  Hereafter we will present the comparison  of results of
   our model and   the observational data.

   As a set of observational
   data,
   the analysis of SNe+CMB data done in the paper~\cite{sahni}
    have been  used. In that paper authors have reconstructed
    the resent history of the Universe on the base of SNe  and CMB
     data in the model-independent way, only modelling DE
      by the hydrodynamical equation of state
\begin{equation}
\label{w} p=w\rho, \quad w=(2q-1)/(3-\Omega_{m}/h^2).
\end{equation}
The cited paper presents two conceptions of the analysis of
observational data: the first of them is the best fit to the data
which uses only the hydrodynamical describing of DE and does not
impose restrictions on the values of $\Omega_{m}$
 and $h_0$, while the second conception follows
 the priority of the  concordance
 $\Lambda$CDM model, so authors of~\cite{sahni}
   put $\Omega_{m}=0.27\pm0.04$ and $h_0=0.71\pm0.06$.
\begin{figure}[ph]
\begin{center}
\includegraphics[width=140 mm,height=6.5 cm]{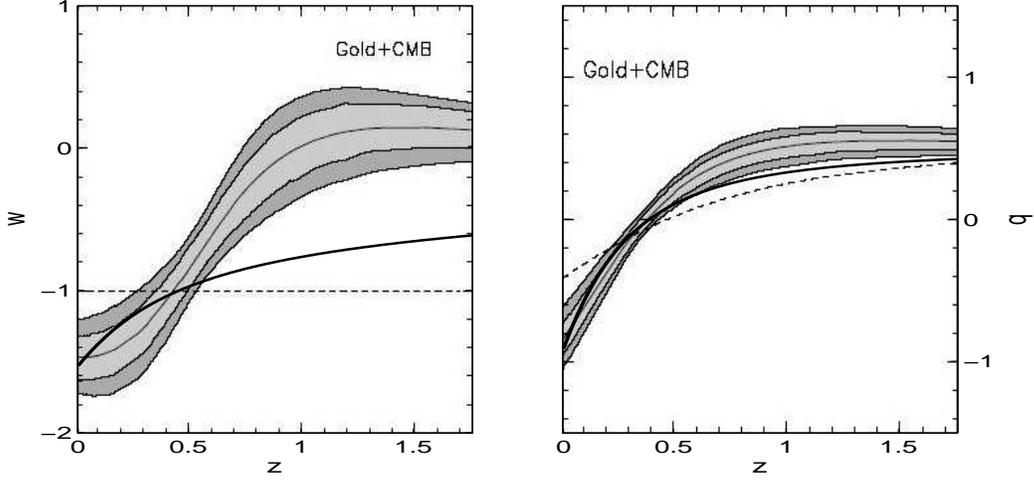}
 \caption{{\footnotesize The comparison of results of the $f(R)\propto \beta R^n$ model
    to the analysis of the
observational data quoted from  the paper~\cite{sahni} by
approbation of the authors. The evolution of the deceleration
parameter with redshift is shown in a right panel and the
variation of the parameter $w$ of equation of state of DE is shown
in a left panel. The best fit of SNe+CMB data with
$\Omega_{m}=0.385,\, h_0=0.60$ is represented by the thin solid
line, the $1\sigma$ and $2\sigma$ confidence levels are
represented by the light and dark grey contours, respectively, and
$\Lambda$CDM is represented by the dashed line. For given
$\Omega_{m}$  and $h$ the $f(R)$-model is defined by parameters
$n=1.2955 $  and $\beta=0.273$ and represented here by the thick
solid line. \label{qw385}}}
\end{center}
 \end{figure}

 We  will give  the comparison of our results with both of them.
Also we notice that the analogy  of the ``hydrodynamical'' DE
(\ref{w}) is not so proper to the  higher order gravity theories,
hence one can expect the comparison over the  values $h$ and $q$
to be more informative than over the value $w$.

It have been found in the paper~\cite{sahni} the best fit values
are: $\Omega_{m}=0.385, h_0=60$. In the  model
     DE evolves in time
strongly enough. In Fig. \ref{qw385} one can see results (the
evolution of the deceleration parameter and  DE equation of state
parameter with redshift) for analysis of paper~\cite{sahni}: the
best fit  is represented by the thin solid line,  the $1\sigma$
and $2\sigma$ confidence levels are represented by the light and
dark grey contours, respectively, and $\Lambda$CDM is represented
by the dashed line. Also for given $\Omega_{m}$ and $h_0$ the
results of the $f(R)$-model with $ n= 1.2955 \,(\delta n=0.0002)$
and $\beta=0.273$ for the ``geometrical equation of state''
parameter $w=(y-ay')/3(y-\Omega_{m}\,a)$ and deceleration
parameter $q$ are presented by the thick solid lines. In given
$f(R)$-model the age of the Universe is $14.9$ Hyr, the
deceleration parameter is $q_0=-0.91$ at present and  the
transition to acceleration occurs at $z=0.38$. Similarly to
results of~\cite{sahni}, the $w_{DE} <1$ at lower redshifts
($w_{DE0}= -1.53$), however , the evolution of equation of state
of ``geometrical DE'' is more weak contrary to the results
of~\cite{sahni}.

 \begin{figure}[pt]
\begin{center}
\includegraphics[width=140 mm,height=6.5 cm]{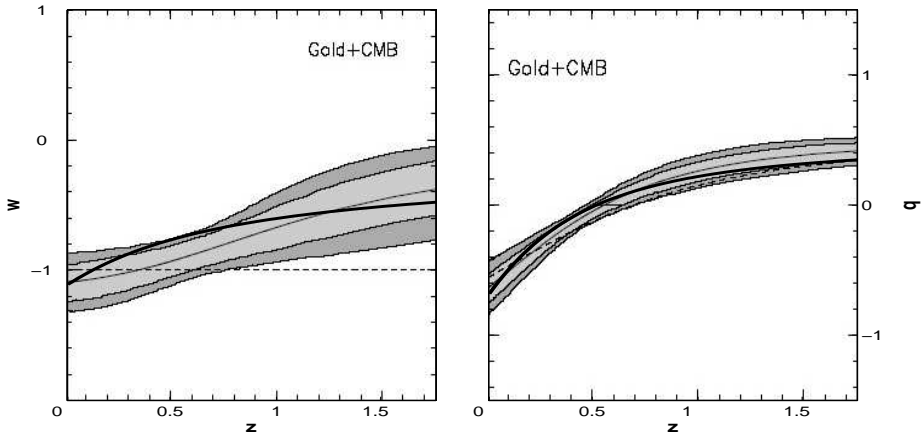}
 \caption{{\footnotesize The comparison of results of $f(R)\propto
R^n$ model with $n=1.296$ and $\beta=0.467$ (the thick solid line)
to results of analysis of SNe+CMB data with $\Omega_{m}=0.29$ done
in~\cite{sahni}. The evolution of the deceleration parameter with
redshift is shown  in a right panel and variation of equation of
state of DE is shown in a left panel. The best fit of SNe+CMB data
in such case is represented by the thin solid line, the $1\sigma$
and $2\sigma$ confidence levels are represented by the light and
dark grey contours, respectively, and $\Lambda$CDM is represented
by the dashed line.
 \label{star29}}}
 \end{center}
 \end{figure}
 However, if strong priors have been
imposed on $\Omega_{m}$ and $h_0$ (i.e. the $\Lambda$CDM model
priors: $\Omega_{m}=0.27\pm0.04$ and $h_0=0.71\pm0.06$), the
evolution of DE is extremely weak and in good agreement with the
$\Lambda$CDM model. The  best fit in the case is
$\Omega_{m}=0.29$. One can see good enough coincidence of our
model and their analysis for parameters of the model $ n=
1.296\,(\delta
  n=0.001)$
and $\beta=0.467$ in figure \ref{star29}. The deceleration
parameter $q_0=-0.683$ at present and the  deceleration was
changed by the acceleration at $z=0.51$, ($q_0=-0.63\pm0.12$ and
$z=0.57\pm0.07$ in the paper~\cite{sahni}). The age of the
Universe in this case is 13.6 Hyr.

\begin{figure}[ph]
\begin{center}
\includegraphics[width=140 mm,height=6.5 cm]{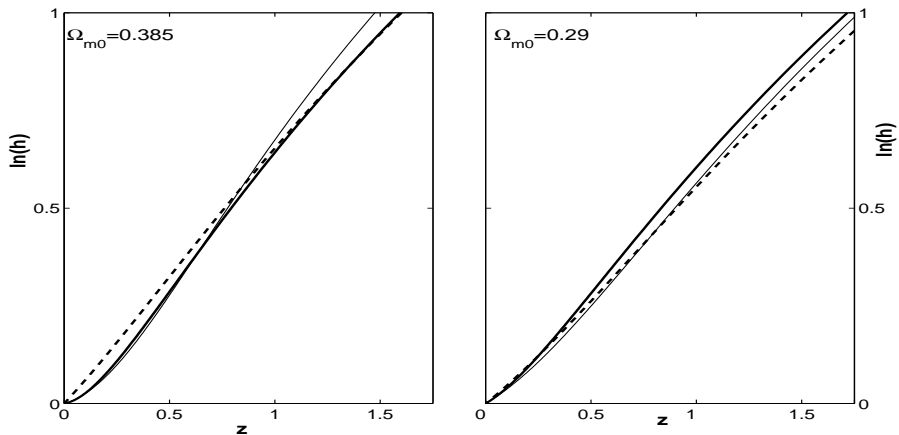}
 \caption{{\footnotesize
 The evolution of the Hubble parameter with
redshift. The $f(R)$-model and the $\Lambda$CDM model are shown by
the thick solid and  the dashed lines, while the best fit of the
observational data according to~\cite{sahni} is represented by the
thin solid line. The case for $\Omega_{m}=0.385,\; h_0=0.6$ is
shown in the left panel, and the case $\Omega_{m}=0.29,\;
h_0=0.68$ is shown in the right panel.  \label{h385_29}}}
\end{center}
 \end{figure}

In Fig. \ref{h385_29}  the comparison of the evolution of the
Hubble parameter for $f(R)$-model to $\Lambda$CDM model and the
best fit of the observational data analysis is presented. One can
see the results  for the cases with $\Omega_{m}=0.385$ and
$\Omega_{m}=0.29$ in a left  and a right panel, respectively.

Thus, the $f(R)\propto \beta R^n$-model with  parameters $\beta$
and $n$ chosen according to the principles mentioned in
Introduction describes  the evolution of the Universe
 quite corresponding to the SNe+CMB data.

\section{Acknowledgements}
V.G is grateful to A. Starobinsky for constructive criticism of
 the paper \cite{fol} which has stimulated  the given investigation.
 Authors also thank A. Nusser, H. Kleinert, and V. Folomeev for helpful
 discussions.

\end{document}